# Cover Story Management

Frederic Cuppens
ONERA/CERT
2 Avenue E. Belin
31055, Toulouse Cedex
France
email: cuppens@cert.fr

Alban Gabillon
Université de Toulon et du Var
SIMM
B.P.132, 83 957, La Garde cedex.
France
email : gabillon@univ-tln.fr

**Abstract**

In a multilevel database, cover stories are usually managed using the ambiguous technique of polyinstantiation. In this paper, we define a new technique to manage cover stories and propose a formal representation of a multilevel database containing cover stories. Our model aims to be a generic model, that is, it can be interpreted for any kind of database (e.g. relational, object-oriented etc). We then consider the problem of updating a multilevel database containing cover stories managed with our technique.

## 1. Introduction

The work presented in this paper applies to the context of multilevel security policies. In a multilevel security policy every piece of information is associated with a *classification* level, and every agent is associated with a *clearance* level. Classification and clearance levels are taken from a set of security levels. This set is partially ordered and forms a lattice. The partial ordering relation is called the *dominance* relation and is denoted by ≥. The *confidentiality property* in the multilevel security policy states that an agent can only know a given piece of information if the clearance level of this agent dominates (that is, is higher than or equal to) the classification level associated with the piece of information.

Currently, database security is an active field of research. Several theoretical models have been suggested for multilevel databases (see, for instance, [DLSSH88,KTT89,HOST90,JK90, Lun90,SW92,ML92,QL96, SC98,JVPJ99]). Most of these models offer the possibility to insert *cover stories* in the multilevel database. Cover stories are lies introduced in the multilevel database in order to protect some existing higher classified data. In most of these models, cover stories are associated with the concept of *polyinstantiation*. Polyinstantiation occurs when different tuples with the same key, each at a different classification level, are allowed. Low level tuples are then interpreted as cover stories. Unfortunately, the concept of polyinstantiation is not satisfactory since it is not free of semantic ambiguities.

This is why we propose in [CG99] a formal model for multilevel databases which do not use polyinstantiation. This model can be interpreted for any type of database (e.g. object-oriented database, relational database etc). Our model offers the possibility to insert *cover stories* into the multilevel database. In our model, we explicitly state which data are the cover stories. We claim that this approach is free of ambiguity and is therefore better than using polyinstantiation. In [CG99], we define a *secure* multilevel database as being a multilevel database free of illegal inference channels, that is, we say that a multilevel database is secure if low level users cannot deduce some high classified data using the low classified data represented in the database. However, in [CG99] we make some restrictive assumptions on the form of the data which can be contained in the database. For instance, we assume that the database cannot contain some disjunctive integrity constraints. Moreover, in [CG99], we do not address the difficult problem of updating a multilevel database containing cover stories. This problem can be summarized as follows: how to guarantee that after an update the multilevel database is still consistent and secure?

The contribution of this paper is the following:

- We give a more general definition of what a secure multilevel database is. With this new definition we do not need to make any restrictive assumption on the form of the data which can be contained in the database

- We propose an algorithm which aims to restore the consistency and the security property of the multilevel database after an update has been performed. This algorithm is *automatically* initiated by the DBMS after each committed transaction.

Note however that in [CG99], we deal with the use of the special "*Restricted*" value (first suggested by Sandhu and Jajodia [SJ92]) as an alternative for cover stories. In this paper, we have decided to focus on cover stories, therefore we have not mentioned this possibility (see [CG99] for the semantics of this special "*Restricted*" value).

The remainder of this paper is organized as follows. Section 2 introduces, in an informal way, the concept of cover story and outlines the problems related to polyinstantiation. Section 3 proposes a model for a multilevel database containing cover stories. Section 4 describes how cover stories can be practically managed and how the consistency and the security property of the database can be maintained after an update has occurred. Section 5 concludes this paper.

## 2. Cover Stories and Polyinstantiation

We shall use the following two lattices throughout the paper:

$U\_S$:  $\{(U)\text{nclassified},(S)\text{ecret}\}$ with $S > U$

$U\_C1\_C2\_S$:  $\{(U)\text{nclassified},(C_1)\text{onfidential1},(C_2)\text{onfidential2},(S)\text{ecret}\}$ with $S>C_1$, $S>C_2$, $C_1>U$, $C_2>U$ and $C_1<>C_2$ (which means $C_1$ and $C_2$ are not comparable)

Intuitively, a cover story is a lie inserted in a multilevel database in order to hide the existence of a sensitive data.

Let us use $U\_S$ and let us consider a multilevel database containing the following fact:

*Salary*(Dupont,2000)    which reads "Dupont's salary is 2000"

Let us assume this fact is classified at the Secret level:

$[Salary(\text{Dupont},2000)]_S$    which reads "the fact that Dupont's salary is 2000 is Secret"

Let us assume an Unclassified user *u* queries the database and asks "What is Dupont's salary?". Obviously, the database cannot answer "Dupont's salary is 2000" since this is a Secret fact. Let us assume that the database answers "You are not allowed to know this data". From this answer, *u* deduces that the salary of Dupont is Secret[1]. In other words, *u* deduces the following sentence:

$\exists x,[Salary(\text{Dupont},x)]_S$    which reads "there exists a Secret salary for Dupont"

However, the Security Administrator (SA) might consider this sentence to be itself Secret that is:

$[\exists x,[Salary(\text{Dupont},x)]_S]_S$    which reads "the fact that there exists a Secret salary for Dupont, is itself Secret"

In this case, the database cannot answer "You are not allowed to know this data"[2]. As a solution the SA may insert a salary at the Unclassified level in the multilevel database.

$[Salary(\text{Dupont},1500)]_U$

---

[1] $U\_S$ contains only two security levels {Unclassified, Secret}. If we consider more than two security levels then *u* would only learn that Dupont's salary is not Unclassified.

[2] Note that, if the SA had considered that the sentence "there exists a Secret salary for Dupont" was Unclassified (that is $[\exists x,[Salary(\text{Dupont},x)]_S]_U$) then the answer "You are not allowed to know this data" would have been acceptable.

Now, there are two salaries for Dupont in our multilevel database: one Unclassified, 1500 and one Secret, 2000. In the literature, the salary of Dupont is referred to as a *polyinstantiated* set of two classified values.

*Polyinstantiation* was first used in the SeaView project [DLSSH88]. With this technique different tuples may have the same primary key provided each of them has a different classification level. Low-classified tuples are implicitly interpreted as cover stories.

Therefore, according to polyinstantiation, 1500 is interpreted as a cover story that is, a lie which is used to hide from Unclassified users the existence of a more sensitive salary for Dupont. 2000 is interpreted as the Dupont's real salary. If $u$ asks "What is Dupont's salary?" then the database answers "Dupont's salary is 1500". Therefore, $u$ cannot deduce the existence of a more sensitive salary. If a Secret user expresses the same query then the database answers "Dupont's salary is 2000" and possibly informs the Secret user about the existence of a cover story at the Unclassified level.

Despite the broad use of polyinstantiation by the community working on multilevel databases, we claim that this technique is not appropriate to cover story management. Our aim, now, is to describe some of the problems which are related to the use of polyinstantiation.

*1. Ambiguous interpretation of data*

Let us consider our previous multilevel database containing the following two classified facts:

$[Salary(\text{Dupont}, 1500)]_U$

$[Salary(\text{Dupont}, 2000)]_S$

With the polyinstantiation technique, 1500 is interpreted as a cover story whereas 2000 is interpreted as the Dupont's real salary. This interpretation implicitly suggests that there exists an integrity constraint stating that the salary of Dupont is unique. However, if the database does not contain such an integrity constraint then this database could also represent the fact that Dupont has *two* salaries: one Unclassified, 1500, and one Secret, 2000. In this latter case a Secret user querying the database and asking "What is Dupont's salary?" should be answered as "2000 and 1500" and not as "2000" only.

*2. Poor expressive power of the security policy*

Polyinstantiation allows us to introduce cover stories which are "attribute values" only (e.g. the salary attribute of the entity Dupont). It does not provide us with any means to introduce an "entity" cover story. For example, we cannot introduce a lie saying that another person, Durand, is an employee.

*3. Polyinstantiation does not work in case of a partial order on the set of security levels*

Let us use *U_C1_C2_S* and let us consider a multilevel database containing the following two classified facts:

$[Salary(\text{Dupont}, 1500)]_{C_1}$

$[Salary(\text{Dupont}, 1500)]_{C_2}$

Let us assume we have an integrity constraint saying that Dupont's salary is unique. At least, one of these two salaries is then a lie. However, polyinstantiation does not provide us with any means to know which of these two salaries is a cover story.

In the remaining of this paper, our objective is to define a new technique for managing cover stories. We show that our technique is free of semantic ambiguities and it offers a security policy with a great expressive power.

## 3. Multilevel database

### 3.1 Classified facts and integrity constraints

Mathematical logic has often been used to formalize databases. In [CG99] we define a complete logical theory which represents a multilevel database. In this paper, in order not to repeat what was written in [CG99] and to highlight our new cover story management technique, we shall only mention the relevant characteristics of our model without going into formal details.

> We consider that a multilevel database $DB$ contains a set $F$ of *classified atomic facts* and a set $IC$ of *classified integrity constraints*.
>
> For the sake of simplicity we also assume that integrity constraints are all classified at the lowest level $U$.

Example:

Let us use $U\_S$.

$F = \{$    $[Employee(\text{Dupont})]_U$    which reads " Dupont is an employee". This fact is $U$.

       $[Salary(\text{Dupont},2000)]_S \}$    which reads "Dupont'salary is 2000". This fact is $S$.

$IC = \{$    $[\forall x \forall y, Salary(x,y) \rightarrow Employee(x)]_U$                                  (1)

       which reads "Anybody who has a salary is an employee". This constraint is $U$.

       $[\forall x, Employee(x) \rightarrow \exists y, Salary(x,y)]_U$                                      (2)

       which reads "each employee has a salary". This constraint is $U$.

Formulae of $F \cup IC$ are axioms of our logical theory.

We also extend our theory with the following axiom schema ($p$ denotes an atomic fact or an integrity constraint. $l$ denotes a security level):

$$\forall l \forall l', [p]_l \wedge [p]_{l'} \wedge l \leq l' \rightarrow l = l' \quad (3)$$

This axiom schema says that a fact or an integrity constraint cannot be associated with two distinct comparable security levels. This means that if $p$ is classified with two levels $l$ and $l'$ and if $l'$ dominates $l$ then $l$ and $l'$ are the same. This axiom implicitly suggests that $p$ may be associated with more than one security level provided these security levels cannot be compared between each other:

Example :

Let us use $U\_C1\_C2\_S$.

According to axiom schema (3), $[Employee(\text{Dupont})]_U \wedge [Employee(\text{Dupont})]_{C_1}$ is not possible since $C_1 > U$ whereas $[Employee(\text{Dupont})]_{C_1} \wedge [Employee(\text{Dupont})]_{C_2}$ is possible since $C_1 <> C_2$.

## 3.2 Cover Stories

A multilevel database may contain lies, which are cover stories. We have seen in section 2 that polyinstantiation is often suggested as a solution to manage cover stories. However, we have also seen in section 2 that there are several problems related to the use of polyinstantiation. Therefore, the aim of this section is to propose a new technique to manage cover stories.

The central idea of our technique is to explicitly declare a cover story in either a set $CS^F$ (if the cover story is a fact) or in a set $CS^{IC}$ (if the cover story is an integrity constraint).

More precisely, if $p$ is a cover story of the multilevel database then we introduce in the multilevel database (in either $CS^F$ or $CS^{IC}$) a classified data saying that $p$ is a cover story.

Let us have an overview of our technique through the following three examples:

Example 1:

Let us use $U\_S$.

$F = \{$  $[Employee(\text{Dupont})]_U$

$[Salary(\text{Dupont},1500)]_U$

$[Salary(\text{Dupont},2000)]_S \}$

$IC = \{$  $[\forall x \forall y, Salary(x,y) \rightarrow Employee(x)]_U$

$[\forall x, Employee(x) \rightarrow \exists y, Salary(x,y)]_U \}$

$CS^F = \{$  $[Salary(\text{Dupont},1500)]_S^{CS} \}$

(which reads "the fact that the salary of Dupont, 1500, is a cover story, is Secret")

$CS^{IC} = \{\}$

With our new cover story management technique, 1500 is interpreted as a cover story without any ambiguity. If $CS^F$ were empty then 1500 would be interpreted as a second salary for Dupont without any ambiguity. Since cover stories are explicitly declared, there is no ambiguity in the interpretation of the data.

Example 2:

Let us use $U\_S$.

$CS^F = \{$  $[Employee(\text{Durand})]_S^{CS} \}$

$F = \{$  $[Employee(\text{Durand})]_U$

$[Salary(\text{Durand}, 1000)]_U$ }

In this example the fact that Durand is an employee is a lie. On the contrary, the fact that Durand receives a salary of 1000 is not a lie ! One can reasonably guess that the information "Durand is an employee" has been introduced at the Unclassified level in order to justify the fact that Durand receives a salary. This example shows that we can create "entity" cover stories.

Note that this database is not consistent with integrity constraint (1). Indeed, Secret users see a salary for Durand but Durand is not an employee. This contradicts integrity constraint (1).

Consequently, the above example is possible if, either *IC* does not contain integrity constraint (1) or, $CS^{IC}$ contains a fact saying that integrity constraint (1) is a cover story:

$IC = \{$   $[\forall x, Employee(x) \rightarrow \exists y, Salary(x,y)]_U \}$   (*IC* contains integrity constraint (2) only)

$CS^{IC} = \{\}$

Or,

$IC = \{$   $[\forall x \forall y, Salary(x,y) \rightarrow Employee(x)]_U$

  $[\forall x, Employee(x) \rightarrow \exists y, Salary(x,y)]_U \}$   (*IC* contains both (1) and (2))

$CS^{IC} = \{ [\forall x \forall y, Salary(x,y) \rightarrow Employee(x)]_S^{CS} \}$   (integrity constraint (1) is a cover story)

Example 3:

Let us use *U_C1_C2_S*.

$CS^F = \{$   $[Salary(\text{Dupont}, 1500)]_S^{CS} \}$

$F = \{$   $[Employee(\text{Dupont})]_{C_1}$

  $[Employee(\text{Dupont})]_{C_2}$

  $[Salary(\text{Dupont}, 1500)]_{C_1}$

  $[Salary(\text{Dupont}, 2000)]_{C_2} \}$

Secret users are explicitly informed that 1500 at the $C_1$ level is a cover story. Our technique allows us to manage cover stories even if we have a partial order on the set of security levels.

These three examples have given us an overview of the expressive power of a security policy which would manage cover stories using our technique.

### 3.3 Definition of a multilevel database

We define the multilevel database *DB* as follows:

$DB = F \cup IC \cup CS^F \cup CS^{IC}$

Formulae belonging to *DB* are axioms of our logical theory.

Axiom schemata specifying the links between the cover stories and the other data are the following (*p* is either a fact or an integrity constraints):

$$\forall l, [p]_l^{CS} \rightarrow \exists l', l' < l \wedge [p]_{l'} \tag{4}$$

Axiom (4) says that if the fact that *p* is a cover story is classified at level *l* then there exists a level *l'* strictly dominated by *l* such that *p* is classified at level *l'*.

Note that from axioms (4) and (3) we can derive the following theorem:

$$\forall l, [p]_l^{CS} \rightarrow \neg [p]_l \tag{5}$$

Theorem (5) says that if the fact that *p* is a cover story is classified at level *l* then *p* cannot be classified at level *l*.

Proof:

Let us assume that $\exists l, [p]_l^{CS}$

From (4) we can derive that $\exists l', [p]_{l'} \wedge l' < l$

Axiom (3) can be rewritten as $\forall l \forall l', [p]_{l'} \wedge l' < l \rightarrow \neg [p]_l$

Therefore, we can derive that $\neg [p]_l$

Derivation in the multilevel database is based on the classical Modus Ponens inference rule (that is, if *p* is a theorem and if "*p* implies *q*" is a theorem then we can derive that *q* is also a theorem).

We write $DB \vdash p$ which reads *p* is derived from *DB*.

We do not use the *closed world assumption* since the closed world assumption generally leads to inconsistency when it is combined with disjunctions. Negative facts can be either derived from some of the axioms given previously (see for instance the theorem 5) or from *axioms of completion* (see [Rei83]). We did not explicitly mentioned these axioms in the previous examples but they should be considered as implicit. For instance, sets *F* and $CS^F$ of example 1 should be rewritten as follows:

$F = \{ \quad [\forall x, [Employee(x)]_U \leftrightarrow x = \text{Dupont}]_U$

$\quad\quad [\forall x \forall y, [Salary(x,y)]_U \leftrightarrow x = \text{Dupont} \wedge y = 1500]_U$

$\quad\quad [\forall x \forall y, [Salary(x,y)]_S \leftrightarrow x = \text{Dupont} \wedge y = 2000]_S \}$

$CS^F = \{ \quad [\forall x \forall y, [Salary(x,y)]_S^{CS} \leftrightarrow x = \text{Dupont} \wedge y = 1500]_S \}$

### 3.4 Definition of a consistent multilevel database

A multilevel database represents a part of the real world. Facts and integrity constraints which belong to the multilevel database, and which are not cover stories, exist also in the world represented by the multilevel database. This can be expressed by the following sentence:

$$\forall l, [p]_l \land \neg(\exists l', l < l' \land [p]_{l'}^{CS}) \to p \qquad (6)$$

Formula (6) says that if $p$ is classified at level $l$ and if $p$ is not declared as a cover story in the upper levels then $p$ is not a lie, that is, $p$ exists in the real world.

Using formula (6), we can easily derive the world represented by the multilevel database. From $F \cup CS^F$, we can derive $F^{Real}$, a set of facts existing in the real world, and from $IC \cup CS^{IC}$ we can derive $IC^{Real}$, a set of integrity constraints existing in the real world.

Example 4:

Let us use $U\_S$. Let us consider the following multilevel database:

$CS^F = \{ \quad [Salary(\text{Dupont}, 1500)]_S^{CS}$

$\qquad\qquad [Employee(\text{Durand})]_S^{CS} \}$

$F = \{ \quad [Employee(\text{Dupont})]_U$

$\qquad\qquad [Salary(\text{Dupont}, 1500)]_U$

$\qquad\qquad [Salary(\text{Dupont}, 2000)]_S$

$\qquad\qquad [Employee(\text{Durand})]_U$

$\qquad\qquad [Salary(\text{Durand}, 1000)]_U \}$

$CS^{IC} = \{ \quad [\forall x \forall y, Salary(x, y) \to Employee(x)]_S^{CS} \}$

$IC = \{ \quad [\forall x \forall y, Salary(x, y) \to Employee(x)]_U$

$\qquad\qquad [\forall x, Employee(x) \to \exists y, Salary(x, y)]_U \}$

The world represented by this multilevel database is the following:

$F^{Real} = \{ \quad Employee(\text{Dupont})$

$\qquad\qquad Salary(\text{Dupont}, 2000)$

$\qquad\qquad Salary(\text{Durand}, 1000) \}$

$IC^{Real} = \{ \quad \forall x, Employee(x) \to \exists y, Salary(x, y) \}$

---

We say that a multilevel database $DB = F \cup IC \cup CS^F \cup CS^{IC}$ is *consistent* if and only if:

1. $F \cup IC \cup CS^F \cup CS^{IC} \vdash \{3,4\}$ which reads "axiom schemas (3) and (4) can be derived from $DB$".

2. $F^{Real} \vdash IC^{Real}$ which reads "integrity constraints of $IC^{Real}$ can be derived from the set of facts $F^{Real}$"

---

One can easily see that our previous example of multilevel database is consistent.

## 3.5 Definition of a secure multilevel database

A multilevel database might be consistent but not free of illegal inference channels. There is an illegal inference channel when a low level user can deduce some sensitive information from the low classified data.

Example 5:

Let us use *U_S* and let us consider the following multilevel database:

$CS^F = \{\}$

$F = \{ \quad [Employee(\text{Dupont})]_U$

$\quad\quad\quad [Salary(\text{Dupont},2000)]_S \}$

$CS^{IC} = \{\}$

$IC = \{ \quad [\forall x \forall y, Salary(x,y) \rightarrow Employee(x)]_U$

$\quad\quad\quad [\forall x, Employee(x) \rightarrow \exists y, Salary(x,y)]_U \}$

The world represented by this database is the following:

$F^{Real} = \{ \quad Employee(\text{Dupont})$

$\quad\quad\quad Salary(\text{Dupont},2000)\}$

$IC^{Real} = \{ \quad \forall x, Employee(x) \rightarrow \exists y, Salary(x,y)$

$\quad\quad\quad \forall x \forall y, Salary(x,y) \rightarrow Employee(x)\}$

One can easily see that this database is consistent. However, it is not free from illegal inference channels. Indeed, Unclassified users can see both the Unclassified fact *Employee*(Dupont) and the Unclassified integrity constraint $\forall x, Employee(x) \rightarrow \exists y, Salary(x,y)$. By assuming that the database is consistent, these users can learn that Dupont has a salary and that his salary is Secret.

Let $DB_l$ be the view of the multilevel database at level *l*. $DB_l$ contains all the data (facts, integrity constraints and cover stories) which are classified at a level lower than or equal to *l*.

$DB_l = F_l \cup IC_l \cup CS_l^F \cup CS_l^{IC}$

Let $F_l^{Real} \cup IC_l^{Real}$ be the world which can be derived from $DB_l$.

---

We say that a consistent database $DB_l$ is *secure* (or respects the *confidentiality property*) if and only if for all security level *l'* (with $l \geq l'$), $DB_{l'}$ is consistent, that is for all security level *l'*, we have:

1. $F_{l'} \cup IC_{l'} \cup CS_{l'}^F \cup CS_{l'}^{IC} \mid\!\!-\{3,4\}$

2. $F_{l'}^{Real} \mid\!\!- IC_{l'}^{Real}$

---

Note that, saying that *DB* is secure is equivalent to say that $DB_{high}$ is secure since $DB = DB_{high}$. *high* is the security level dominating all the other security levels (recall that the set of security levels forms a lattice).

Our previous multilevel database is not secure because $DB_U$ is not consistent:

$CS_U^F = \{\}$

$F_U = \{ \quad [Employee(\text{Dupont})]_U \}$

$CS_U^{IC} = \{\}$

$IC_U = \{ \quad [\forall x \forall y, Salary(x, y) \rightarrow Employee(x)]_U$

$\quad\quad\quad\quad [\forall x, Employee(x) \rightarrow \exists y, Salary(x, y)]_U \}$

The world represented by $DB_U$ is the following:

$F_U^{Real} = \{ \quad Employee(\text{Dupont}) \}$

$IC_U^{Real} = \{ \quad \forall x, Employee(x) \rightarrow \exists y, Salary(x, y)$

$\quad\quad\quad\quad \forall x \forall y, Salary(x, y) \rightarrow Employee(x) \}$

We see that $F_U^{Real} \not\vdash IC_U^{Real}$. In particular $F_U^{Real} \not\vdash \forall x, Employee(x) \rightarrow \exists y, Salary(x, y)$.

Note that in [CG99], we state two *inference rules* which allow us to derive some *inference control theorems* from the integrity constraints. In [CG99], we say that a multilevel database is secure if it respects these theorems. The definition used in this paper is more general. Indeed, in [CG99] we make some assumptions on the form of integrity constraints we may insert in the database. In particular, we assume we cannot have disjunctive integrity constraints such as:

$\forall x, Human(x) \rightarrow Man(x) \vee Woman(x)$

which reads "every human being is either a man or a woman".

In this paper we do not make any restrictive assumption on the form of integrity constraints.

We could show that if a database is secure regarding the definition given in [CG99] then it is also secure regarding the definition given in this paper.

Note also that "*DB* being secure" means only that the integrity constraints cannot be used to derive protected data. Of course, a low level user might still be able to deduce some high classified data from an external knowledge not represented in the database.

The aim of the next section is to investigate the problem of updating a multilevel database containing cover stories which are handled with our technique.

12## 4. Updating a multilevel database

### 4.1 Single-level Transactions

Let *DB* be a secure multilevel database. We assume *DB* can be updated through single-level transactions only. Before starting a single-level transaction, a user must first choose a *transaction level l*. This transaction level *l* must be dominated by the clearance level of the user. Then, the user may read data at a level dominated by *l* and may insert/update/delete data at a level equal to *l*, this, in accordance with the No-Read-Up and No-Write-Down principles of the Bell & LaPadula model [BL75].

We assume there is no restriction on the data the user may access within the transaction. This means that a user may access data from *F*, *IC*, $CS^F$, $CS^{IC}$.

> We say that the commitment of a transaction of level *l* is accepted if $DB_l$ is secure. It is rejected otherwise.

This principle seems reasonable. A transaction of level *l* being rejected means that $DB_l$ is not secure. Since a user of level *l* is allowed to see the whole $DB_l$ database, rejecting the transaction does not violate the confidentiality property.

On the other hand, if $DB_l$ is secure then the transaction must be accepted and committed, even if the whole *DB* database is not secure. Indeed rejecting the transaction in case $DB_l$ is secure and *DB* not secure would open a covert channel.

Example 6:

Let us use *U_S* and let us consider the database *DB* represented in the example 1 of page 6:

Let us now consider two examples of Unclassified transactions updating *DB*:

1. Let *T* be an Unclassified transaction inserting in *F* the new fact *Employee*(Durand). *T* is rejected since the new $DB_U$ would not be consistent. Indeed there would be a new employee Durand without a salary in $F_U^{Real}$. This would contradict the integrity constraint $\forall x, Employee(x) \rightarrow \exists y, Salary(x, y)$ belonging to $IC_U^{Real}$. Therefore *DB* is not updated.

2. Let *T* be an Unclassified transaction updating the fact [*Salary*(Dupont,1500)]$_U$ into the fact [*Salary*(Dupont,1600)]$_U$. *T* is accepted and committed since the new $DB_U$ is consistent. *DB* becomes the following:

$CS^F = \{$ $\quad [Salary(\text{Dupont},1500)]_S^{CS} \}$

$F = \{$ $\quad [Employee(\text{Dupont})]_U$

$\quad\quad\quad [Salary(\text{Dupont},\mathbf{1600})]_U$

$\quad\quad\quad [Salary(\text{Dupont},2000)]_S \}$

$CS^{IC} = \{\}$

$IC = \{$ $\quad [\forall x \forall y, Salary(x, y) \rightarrow Employee(x)]_U$

$$[\forall x, Employee(x) \rightarrow \exists y, Salary(x, y)]_U \}$$

We can easily see that $DB_U$ is consistent. However, we also see that $DB$ is not consistent (in particular, we see that $DB$ is not consistent with axiom (4). Indeed, there is a fact at level $S$ saying that $Salary$(Dupont,1500) is a cover story whereas it does no longer appear in the unclassified database

Consequently, the main question is, what should be done in such a case to make the whole $DB$ secure again ?

There are several possible answers to this question. We are going to deal with them in the next section of this paper. However, the following preliminary action must be taken in any case:

> If $DB$ is not secure after a transaction has been committed, then a mechanism alerting the Security Administrator (SA) must be triggered.

Of course this mechanism must be triggered through a trusted path and must not be seen by low level users.

### 4.2 Restoring the security property – Basic algorithm

Let us consider a secure database $DB$. Let us assume $DB$ is updated by a transaction $T$. $T$ is committed. Let us assume the new $DB$ is not secure.

According to the principle enounced in the previous section, a mechanism informing the SA has been triggered. The SA may then take any decision that he estimates the best to restore the security property of $DB$. However, the SA is a human being. He cannot react instantly. Consequently, during a certain period of time, $DB$ remains in an non-secure state. This may be unacceptable.

The solution to cope with this problem is to let the DBMS take a decision and restore *immediately* the confidentiality property of $DB$. Of course, the SA might later reverse the operations performed by the DBMS and perform other operations if necessary.

We suggest that the DBMS applies the following algorithm in order to restore the confidentiality property of $DB$:

1. If $DB$ is not consistent with axiom (3) that is, there exists a fact or a constraint $p$ associated with two comparable security levels $l$ and $l'$ with $l < l'$, then $[p]_{l'}$ is deleted.

    $[p]_{l'}$ is deleted since $p$ is disclosed at a level strictly lower than $l'$.

2. If $DB$ is not consistent with axiom (4) that is, there exists a $l$-classified data saying that a fact or a constraint $p$ is a cover story and there is no security level $l'$ (with $l' < l$) protecting $p$, then $[p]_l^{CS}$ is deleted.

    $[p]_l^{CS}$ is deleted since $p$ is not contained in the database at any level strictly lower than $l$.

3. Let $I = \{ p_1, p_2, \ldots, p_n \}$. Each $p_i$ is either a fact belonging to $F$ or an integrity constraints belonging to $IC$.

    $I$ is inconsistent if and only if the integrity constraints belonging to $I$ cannot be derived from the facts belonging to $I$.

*I* is minimal inconsistent if and only if *I* is inconsistent and every strict subset of *I* is consistent.

Let us consider the following rule to restore the database consistency:

If $I = \{p_1, p_2, \ldots, p_n\}$ is a minimal inconsistent set and

if we have $[p_1]_{l_1} \wedge \ldots \wedge [p_n]_{l_n} \wedge l = \text{lub}(l_1, \ldots, l_n)$ [3],

then we have:

$$[p_1]_l^{CS} \vee \ldots \vee [p_n]_l^{CS}$$

This means, at least one of the $p_i$ is a cover story and the fact that it is a cover story is classified at level *l*.

We call this rule the *cs-rule*.

Using this cs-rule the DBMS must attempt to derive some new cover stories.

This algorithm is a general algorithm which may be refined. Step 3 of this algorithm says that if *I* is a minimal inconsistent set then one of the $p_i$ is a cover story. Recall that each $p_i$ is either a fact or an integrity constraint. Recall also that step 3 of the algorithm is used by the DBMS to derive some new cover stories. Knowing this, we could decide that the DBMS should trust the integrity constraints more than the facts. Therefore, we could refine the algorithm with the following step 4:

4. If the DBMS could derive $[p_1]_l^{CS} \vee \ldots \vee [p_n]_l^{CS}$ from step 3 and if $p_1, p_2, \ldots, p_k$ are facts and $p_{k+1}, p_{k+2}, \ldots, p_n$ are integrity constraints with $1 \leq k < n$

   then we have $[p_1]_l^{CS} \vee \ldots \vee [p_k]_l^{CS}$ that is, one of the facts is a cover story.

   This means that the DBMS should consider that integrity constraints belonging to *I* cannot be cover stories.

In the remainder of this paper we shall assume that step 4 is part of the algorithm.

In many cases, our algorithm will enable the DBMS to restore the security property of *DB*. However, in some cases (see example 10 below), the DBMS will only succeed in deriving a sentence like $[p_1]_l^{CS} \vee \ldots \vee [p_k]_l^{CS}$ from *DB* and will not be able to determine which of the $p_i$ is a cover story. If this occurs, then a solution might be to accept that *DB* remains in a non-secure state until the SA intervenes. Before the SA restores the security property of *DB*, the DBMS must provide *l*-level or higher-level users with the sentence $[p_1]_l^{CS} \vee \ldots \vee [p_k]_l^{CS}$, that is, the DBMS must inform these users that *DB* is not secure and that one of the $p_i$ is a cover story. However, avoiding a situation where the DBMS fails in restoring the security property of *DB* is possible, but it requires some complementary mechanisms or principles which are suggested in section 4.2.

Let us consider the following examples:

Example 7:

---

[3] *Lub* = least upper bound

Let us use *U_S* and let us consider the database *DB* represented in the example 1 of page 6. The following constraint is added to the set *IC* of integrity constraint:

$$[\forall x \forall y \forall y', Salary(x, y) \land Salary(x, y') \rightarrow y = y']_U \qquad (7)$$

(which reads "the salary is unique". This constraint is Unclassified)

Let *T* be an Unclassified transaction updating the fact $[Salary(Dupont,1500)]_U$ into the fact $[Salary(Dupont,1600)]_U$. *T* is accepted and committed since the new $DB_U$ is consistent. As a result, the database becomes the following:

$CS^F = \{ \qquad [Salary(Dupont,1500)]_S^{CS} \}$

$F = \{ \qquad [Employee(Dupont)]_U$

$\qquad\qquad [Salary(Dupont,\mathbf{1600})]_U$

$\qquad\qquad [Salary(Dupont,2000)]_S \}$

$CS^{IC} = \{\}$

$IC = \{ \qquad [\forall x \forall y, Salary(x, y) \rightarrow Employee(x)]_U$

$\qquad\qquad [\forall x, Employee(x) \rightarrow \exists y, Salary(x, y)]_U$

$\qquad\qquad [\forall x \forall y \forall y', Salary(x, y) \land Salary(x, y') \rightarrow y = y']_U \}$

*DB* is not consistent. The DBMS must attempt to restore the security property of *DB*:

- According to step 2 of our algorithm the DBMS deletes $[Salary(Dupont,1500)]_S^{CS}$.

- According to step 3 and 4 of our algorithm the DBMS derives

  $[Salary(Dupont,1600)]_S^{CS}$

Proof:

The set, $\{Salary(Dupont,1600), Salary(Dupont,2000),$
$\qquad\qquad \forall x \forall y \forall y', Salary(x, y) \land Salary(x, y') \rightarrow y = y' \}$ is minimal inconsistent.

From $[Salary(Dupont,2000)]_S \land [Salary(Dupont,1600)]_U$ and from the cs-rule defined in step 3 of our algorithm, we derive

$[Salary(Dupont,2000)]_S^{CS} \lor [Salary(Dupont,1600)]_S^{CS}$
$\lor [\forall x \forall y \forall y', Salary(x, y) \land Salary(x, y') \rightarrow y = y']_S^{CS}$

$[Salary(Dupont,2000)]_S^{CS}$ would violate the consistency. *DB* would not be consistent with theorem (5). Thus, we derive:

$[Salary(Dupont,1600)]_S^{CS} \lor [\forall x \forall y \forall y', Salary(x, y) \land Salary(x, y') \rightarrow y = y']_S^{CS}.$

From step 4 of our algorithm we derive

$[Salary(Dupont,1600)]_S^{CS}$

Then *DB* becomes the following:

$CS^F = \{$        $[Salary(Dupont,1600)]_S^{CS} \}$

$F \ = \{$        $[Employee(Dupont)]_U$

               $[Salary(Dupont,1600)]_U$

               $[Salary(Dupont,2000)]_S\}$

$CS^{IC} = \{\}$

$IC = \{$        $[\forall x \forall y, Salary(x,y) \rightarrow Employee(x)]_U$

               $[\forall x, Employee(x) \rightarrow \exists y, Salary(x,y)]_U$

               $[\forall x \forall y \forall y', Salary(x,y) \wedge Salary(x,y') \rightarrow y = y']_U \}$

One can easily see that *DB* is secure. 1600 is a new cover story. The DBMS could restore the confidentiality property of *DB*.

However, recall that the SA has been informed about the update by the trigger defined in section 4.1. The SA might consider that 1600 is not a cover story and is actually the new real Dupont's salary. If it is the case, then the SA must update *DB* accordingly. In other words he must delete both $[Salary(Dupont,1600)]_S^{CS}$ and $[Salary(Dupont,2000)]_S$.

Example 8:

Let us use *U_S* and let us consider the database *DB* represented in the example 1 of page 6. This database is the same as the initial database defined in the previous example. The only difference is that constraint (7) has been removed. Let us consider the same transaction *T* executed at level *U* which updates the fact $[Salary(Dupont,1500)]_U$ into the fact $[Salary(Dupont,1600)]_U$. *T* is accepted and committed since the new $DB_U$ is consistent. Then the database becomes the following:

$CS^F = \{$        $[Salary(Dupont,1500)]_S^{CS} \}$

$F \ = \{$        $[Employee(Dupont)]_U$

               $[Salary(Dupont,\mathbf{1600})]_U$

               $[Salary(Dupont,2000)]_S\}$

$CS^{IC} = \{\}$

$IC = \{$        $[\forall x \forall y, Salary(x,y) \rightarrow Employee(x)]_U$

               $[\forall x, Employee(x) \rightarrow \exists y, Salary(x,y)]_U \ \}$

*DB* is not consistent. The DBMS must attempt to restore the security property of *DB*:

- According to step 2 of our algorithm the DBMS deletes $[Salary(Dupont,1500)]_S^{CS}$.

Then *DB* becomes the following:

$CS^F = \{\}$

$F = \{\quad [Employee(\text{Dupont})]_U$

$\quad\quad\quad [Salary(\text{Dupont},1600)]_U$

$\quad\quad\quad [Salary(\text{Dupont},2000)]_S\}$

$CS^{IC} = \{\}$

$IC = \{\quad [\forall x \forall y, Salary(x,y) \to Employee(x)]_U$

$\quad\quad\quad [\forall x, Employee(x) \to \exists y, Salary(x,y)]_U \}$

*DB* is secure. 1600 is now interpreted as Dupont's second salary. The DBMS could restore the security property of *DB*.

Note that the SA has been informed about the update. Therefore, he may later update *DB* and insert the fact $[Salary(Dupont,1600)]_S^{CS}$ if he estimates that 1600 is actually a cover story.

Example 9:

Let us use *U_S* and let us consider the database *DB* represented in the example 1 of page 6:

Let us consider a transaction *T* executed at level *U* which updates the fact $[Salary(\text{Dupont},1500)]_U$ into the fact $[Salary(\text{Dupont},2000)]_U$. *T* is accepted and committed since the new $DB_v$ is consistent. The database becomes the following:

$CS^F = \{\quad [Salary(Dupont,1500)]_S^{CS} \}$

$F = \{\quad [Employee(\text{Dupont})]_U$

$\quad\quad\quad [Salary(\text{Dupont},\mathbf{2000})]_U$

$\quad\quad\quad [Salary(\text{Dupont},2000)]_S\}$

$CS^{IC} = \{\}$

$IC = \{\quad [\forall x \forall y, Salary(x,y) \to Employee(x)]_U$

$\quad\quad\quad [\forall x, Employee(x) \to \exists y, Salary(x,y)]_U \}$

*DB* is not consistent. The DBMS must attempt to restore the confidentiality property of *DB*:

- According to step 1 of our algorithm the DBMS deletes $[Salary(\text{Dupont},2000)]_S$.

- According to step 2 of our algorithm the DBMS deletes $[Salary(Dupont,1500)]_S^{CS}$.

Then *DB* becomes the following:

$CS^F = \{\}$

$F = \{$         $[Employee(\text{Dupont})]_U$

               $[Salary(\text{Dupont},2000)]_U \}$

$CS^{IC} = \{\}$

$IC = \{$        $[\forall x \forall y, Salary(x,y) \rightarrow Employee(x)]_U$

               $[\forall x, Employee(x) \rightarrow \exists y, Salary(x,y)]_U \}$

*DB* is now secure. We see that the DBMS has actually *downgraded* the value 2000. We claim that this is the most appropriate response to the insertion of the value 2000 at the level *U*, regardless of who initiated the transaction *T*:

- Let us assume that the user who initiated *T* is an Unclassified user. Most probably this user updated the value 1500 into the value 2000 because he *learned* that the real salary of Dupont was 2000. This disclosure means that a violation of the confidentiality property has occurred. Therefore, there is no reason to keep the value 2000 Secret in the database. Of course, the SA who was informed about the update must now investigate and try to discover how the confidentiality property has been violated.

- Let us assume that the user who initiated *T* is a Secret user operating at the level *U*. Our interpretation is that this user updated the value 1500 into the value 2000 because he *wanted* to downgrade the value 2000.

Example 10:

Let us use *U_C1_C2_S* and let us consider the following database *DB*:

$CS^F = \{$       $[Salary(Dupont,1500)]_S^{CS} \}$

$F = \{$         $[Employee(\text{Dupont})]_{C_1}$

               $[Employee(\text{Dupont})]_{C_2}$

               $[Salary(Dupont,1500)]_{C_1}$

               $[Salary(Dupont,2000)]_{C_2} \}$

$CS^{IC} = \{\}$

$IC = \{$        $[\forall x \forall y, Salary(x,y) \rightarrow Employee(x)]_U$

               $[\forall x, Employee(x) \rightarrow \exists y, Salary(x,y)]_U$

               $[\forall x \forall y \forall y', Salary(x,y) \wedge Salary(x,y') \rightarrow y = y']_U \}$

Let us consider a transaction *T* executed at level $C_1$ which updates the fact $[Salary(Dupont,1500)]_{C_1}$ into the fact $[Salary(Dupont,1600)]_{C_1}$. $DB_{C_1}$ becomes the following:

$CS^F_{C_1} = \{\}$

$$F_{C_1} = \{ \quad [Employee(\text{Dupont})]_{C_1}$$
$$[Salary(\text{Dupont}, 1600)]_{C_1} \}$$

$$CS_{C_1}^{IC} = \{\}$$

$$IC_{C_1} = \{ \quad [\forall x \forall y, Salary(x, y) \rightarrow Employee(x)]_U$$
$$[\forall x, Employee(x) \rightarrow \exists y, Salary(x, y)]_U$$
$$[\forall x \forall y \forall y', Salary(x, y) \wedge Salary(x, y') \rightarrow y = y']_U \}$$

One can easily see that $DB_{C_1}$ is secure. $T$ is then accepted and committed. $DB$ becomes the following:

$$CS^F = \{ \quad [Salary(Dupont, 1500)]_S^{CS} \}$$

$$F = \{ \quad [Employee(\text{Dupont})]_{C_1}$$
$$[Employee(\text{Dupont})]_{C_2}$$
$$[Salary(\text{Dupont}, \mathbf{1600})]_{C_1}$$
$$[Salary(\text{Dupont}, 2000)]_{C_2} \}$$

$$CS^{IC} = \{\}$$

$$IC = \{ \quad [\forall x \forall y, Salary(x, y) \rightarrow Employee(x)]_U$$
$$[\forall x, Employee(x) \rightarrow \exists y, Salary(x, y)]_U$$
$$[\forall x \forall y \forall y', Salary(x, y) \wedge Salary(x, y') \rightarrow y = y']_U \}$$

$DB$ is not consistent. The DBMS must attempt to restore the security property of $DB$:

- According to step 2 of our algorithm the DBMS deletes $[Salary(Dupont, 1500)]_S^{CS}$.

- According to step 3 and 4 of our algorithm the DBMS derives

    $[Salary(\text{Dupont}, 2000)]_S^{CS} \vee [Salary(\text{Dupont}, 1600)]_S^{CS}$

    Proof:

    The set, $\{Salary(\text{Dupont}, 1600), Salary(\text{Dupont}, 2000),$
    $\forall x \forall y \forall y', Salary(x, y) \wedge Salary(x, y') \rightarrow y = y' \}$ is minimal inconsistent.

    From $[Salary(\text{Dupont}, 2000)]_{C_2} \wedge [Salary(\text{Dupont}, 1600)]_{C_1}$ and from the cs-rule defined in step 3 of our algorithm, we derive:

$$[Salary(\text{Dupont},2000)]_S^{CS} \vee [Salary(\text{Dupont},1600)]_S^{CS}$$
$$\vee [\forall x \forall y \forall y', Salary(x,y) \wedge Salary(x,y') \rightarrow y=y']_S^{CS}$$

From step 4 of our algorithm, we derive

$$[Salary(\text{Dupont},2000)]_S^{CS} \vee [Salary(\text{Dupont},1600)]_S^{CS}$$

However the DBMS cannot derive which of the two values is a cover story. The simplest solution is to wait for the intervention of the SA.

DB becomes the following:

$CS^F = \{\}$

$F = \{$      $[Employee(\text{Dupont})]_{C_1}$

$[Employee(\text{Dupont})]_{C_2}$

$[Salary(\text{Dupont},\mathbf{1600})]_{C_1}$

$[Salary(\text{Dupont},2000)]_{C_2}$ $\}$

$CS^{IC} = \{\}$

$IC = \{$      $[\forall x \forall y, Salary(x,y) \rightarrow Employee(x)]_U$

$[\forall x, Employee(x) \rightarrow \exists y, Salary(x,y)]_U$

$[\forall x \forall y \forall y', Salary(x,y) \wedge Salary(x,y') \rightarrow y=y']_U$ $\}$

*DB* is not consistent. Until the SA updates the database and restores the consistency of *DB*, Secret users will be informed that either 1600 or 2000 is a cover story.

Solutions to avoid such a situation exist. We briefly present some of them in the next section.

### 4.3    Restoring the security property – Toward an Extended algorithm

The algorithm given in the previous section is a basic algorithm. Example 10 of the previous section has shown us that our algorithm should be extended in order to cope with the situation where the algorithm derives a sentence like $[p_1]_l^{CS} \vee \ldots \vee [p_k]_l^{CS}$ from the database but cannot determine which of the $p_i$ is a cover story. Let us come back to example 10 and let us present some possible extensions for our algorithm.

- The first possibility is to make a non-deterministic choice between the conflicting data. For instance, in example 10, the DBMS can restore the database consistency by non-deterministically deciding that 1600 or 2000 is a cover story. It is the simplest solution and it can apply in every situation. The main drawback is that it is completely arbitrary.

- The SA (or any authorized user) may associate the initial cover story 1500 with a Secret trigger on update. This trigger would automatically insert, through a trusted path, a fact saying that the new updated value at the level $C_1$ is a cover story.

  The database would automatically become as follows:

  $CS^F = \{ \ [Salary(\text{Dupont}, 1600)]_S^{CS} \}$

  $F = \{ \ [Employee(\text{Dupont})]_{C_1}$

  $[Employee(\text{Dupont})]_{C_2}$

  $[Salary(\text{Dupont}, \mathbf{1600})]_{C_1}$

  $[Salary(\text{Dupont}, 2000)]_{C_2} \ \}$

  $CS^{IC} = \{ \}$

  $IC = \{ \ [\forall x \forall y, Salary(x, y) \rightarrow Employee(x)]_U$

  $[\forall x, Employee(x) \rightarrow \exists y, Salary(x, y)]_U$

  $[\forall x \forall y \forall y', Salary(x, y) \wedge Salary(x, y') \rightarrow y = y']_U \ \}$

  $DB$ is now secure. 1600 is a new cover story.

- Another possible solution, which was first suggested in [HOST90] for multilevel relational database and which was further refined in [BCGY93][BCGY94][CG97] in the context of multilevel object-oriented databases, is to use special *pointer* values from higher to lower security levels. For instance, in our previous example, one may have the following initial database:

  $CS^F = \{ \ [Salary(Dupont, \boldsymbol{C_1.Salary.Dupont})]_S^{CS} \}$

  $F = \{ \ [Employee(\text{Dupont})]_{C_1}$

  $[Employee(\text{Dupont})]_{C_2}$

  $[Salary(\text{Dupont}, 1500)]_{C_1}$

  $[Salary(\text{Dupont}, 2000)]_{C_2} \ \}$

  $CS^{IC} = \{ \}$

  $IC = \{ \ [\forall x \forall y, Salary(x, y) \rightarrow Employee(x)]_U$

  $[\forall x, Employee(x) \rightarrow \exists y, Salary(x, y)]_U$

  $[\forall x \forall y \forall y', Salary(x, y) \wedge Salary(x, y') \rightarrow y = y']_U \ \}$

Above expression $C_1.Salary.Dupont$ can be seen as a pointer to the value of Dupont's salary at level $C_1$. Fact $[Salary(Dupont, C_1.Salary.Dupont)]_S^{CS}$ says that Dupont's salary at level $C_1$ is a cover story. Updating the value 1500 into the value 1600 will not affect the consistency of *DB*. Consequently, after the update, 1600 will automatically become a cover story.

- Finally, [CC95] suggests that it is often possible to restore the database consistency by making an implicit choice between incomparable security levels. In particular, this implicit choice may be based on need to know requirements and may depend on the type of information to be updated. For instance, regarding the salary, the SA may specify that level $C_2$ has priority over level $C_1$. Therefore, in our previous example, the conclusion would be to consider that 2000 is Dupont's salary and 1600 is a cover story.

However, there are situations where none of the above solutions, except the first one - namely the non-deterministic choice - applies. The following example provides an example of such a situation.

Let us use *U_S* and let us consider the following database *DB*:

$CS^F = \{\}$

$F = \{ \quad [Employee(Dupont)]_U$

$\quad\quad\quad [Salary(Dupont, 1500)]_U$

$\quad\quad\quad [Salary(Dupont, 2000)]_S \}$

$CS^{IC} = \{\}$

$IC = \{ \quad [\forall x \forall y, Salary(x, y) \rightarrow Employee(x)]_U$

$\quad\quad\quad [\forall x, Employee(x) \rightarrow \exists y, Salary(x, y)]_U$

$$\left[ \begin{array}{l} \forall x \forall y \forall y' \forall y'', Salary(x, y) \wedge Salary(x, y') \wedge Salary(x, y'') \\ \quad \rightarrow (y = y') \vee (y = y'') \vee (y' = y'') \end{array} \right]_U \}$$

which reads "nobody can have more than *two* salaries"

Note that Dupont has two salaries, one Unclassified 1500, and one Secret 2000. Neither of these salaries is a cover story.

Let us consider a transaction *T* executed at level *U* which inserts the fact $[Salary(Dupont, 1600)]_U$. *T* is accepted and committed since the new $DB_U$ is consistent (There are now two salaries at the Unclassified level. This does not violate the integrity constraint which says that nobody can have more than two salaries). The database becomes the following:

$CS^F = \{\}$

$F = \{ \quad [Employee(Dupont)]_U$

$\quad\quad\quad [Salary(Dupont, 1500)]_U$

$\quad\quad\quad [Salary(Dupont, 1600)]_U$

$\quad\quad\quad [Salary(Dupont, 2000)]_S \}$

$CS^{IC} = \{\}$

$IC = \{$ $\quad [\forall x \forall y, Salary(x,y) \to Employee(x)]_U$

$\quad\quad\quad [\forall x, Employee(x) \to \exists y, Salary(x,y)]_U$

$\quad\quad\quad \begin{bmatrix} \forall x \forall y \forall y' \forall y'', Salary(x,y) \wedge Salary(x,y') \wedge Salary(x,y'') \\ \to (y=y') \vee (y=y'') \vee (y'=y'') \end{bmatrix}_U \}$

However, *DB* is not consistent. Indeed, we now have *three* salaries for Dupont.

In a similar way as we did before, we could show that our algorithm derives the following sentence:

$[Salary(\text{Dupont},1500)]_S^{CS} \vee [Salary(\text{Dupont},1600)]_S^{CS}$

that is one of the value 1500 or 1600 is a cover story.

In such a case, in order to restore the consistency of *DB*, the best solution seems to let the DBMS choose in a non-deterministic way which of the two values is a cover story. For instance the DBMS may choose that 1500 is a cover story. *DB* becomes then the following:

$CS^F = \{ \quad [Salary(Dupont,1500)]_S^{CS} \}$

$F = \{ \quad [Employee(\text{Dupont})]_U$

$\quad\quad\quad [Salary(\text{Dupont},1500)]_U$

$\quad\quad\quad [Salary(\text{Dupont},1600)]_U$

$\quad\quad\quad [Salary(\text{Dupont},2000)]_S \}$

$CS^{IC} = \{\}$

$IC = \{$ $\quad [\forall x \forall y, Salary(x,y) \to Employee(x)]_U$

$\quad\quad\quad [\forall x, Employee(x) \to \exists y, Salary(x,y)]_U$

$\quad\quad\quad \begin{bmatrix} \forall x \forall y \forall y' \forall y'', Salary(x,y) \wedge Salary(x,y') \wedge Salary(x,y'') \\ \to (y=y') \vee (y=y'') \vee (y'=y'') \end{bmatrix}_U \}$

*DB* is now secure. 1600 and 2000 are the two Dupont' salaries. 1500 is a cover story.

Examples 7 to 9 of section 4.2 have shown us how our basic algorithm may enable the DBMS to maintain the security property of a multilevel database after an update has occurred. However, we have seen through example 10 that this basic algorithm cannot face all situations and must be extended with some complementary principles, some of which have just been presented in this section.

24## 5. Conclusion

There are situations where it is necessary to insert covert stories in a multilevel database. In section 2, we showed that these situations occur when the security administrator decides to hide the existence of high-classified information from low-classified users. We also showed the limitations of the polyinstantiation technique and suggested a new approach to manage cover stories.

This approach is based on explicit management of cover stories. It enables us to derive a consistent view of the real world for each security level. Moreover, a user cleared at a given security level $l$ is allowed to know all the cover stories having a classification level strictly dominated by $l$.

As mentioned in section 3.5, this new approach is fully compatible with the one suggested by the authors in [CG99]. However, our approach is more general since it does not make any restriction on the form of integrity constraints we may insert in the database.

There are other formal multilevel database security models which do not use the polyinstantiation technique. Smith and Winslett [SW92] were the first to propose such a model. In their model, a multilevel database is represented by a collection of belief sets associated with each security level. This approach was analyzed and criticized in [Cup96]. In particular, it is shown that there is no link between the belief set associated with a given security level $l$ and the belief sets at security levels strictly dominated by $l$. The conclusion of [Cup96] is that [SW92] is not really a model for a multilevel database but rather a model for a "system-high" database.

Recently, [SC98] and [JVPJ99] suggested several improvements to [SW92]. Links can be created between a database at a given security level $l$ and the ones which are at a level strictly dominated by $l$. As a matter of fact, a user cleared at a level higher than or equal to $l$ can specify that a fact classified at a level strictly dominated by $l$ is accepted in the database at level $l$. This is a cautious approach that is efficient to avoid conflicts due to the existence of cover stories at classification levels strictly dominated by $l$. However, we argue that such a manual management of information flow from low to high classification levels is impractical in real applications.

Therefore, we guess that our approach is more realistic. We suggest an explicit management of cover stories so that any fact whose classification is strictly dominated by $l$ is implicitly (and automatically) accepted at level $l$ as far as it is not declared as a cover story at level $l$. Moreover, this explicit management makes it possible to eliminate the problems of information choice, for instance in the case of partially ordered set of security levels.

Finally, we have specified rules to control database updates and deletions, especially when a cover story is modified. From a more practical point of view, the data specifying that a given fact is actually a cover story may be stored in specific relations distinct from the ones used to manage the multilevel database. For instance, if $R$ is a multilevel relation, then one can create a relation $R^{CS}$ to store the cover stories associated with relation $R$. We guess that this overall approach is the most suitable one for future implementations.